\pdfoutput=1 
\documentclass[
  aps,%
  prb,%
 twocolumn,%
 groupedaddress,%
 superscriptaddress,%
  showpacs,%
 letterpaper,%
 amsfonts,%
 footinbib,%
 floatfix,%
]{revtex4-2}

\usepackage{hyperref}
\hypersetup{
    colorlinks=true,
    citecolor=blue,
    linkcolor=blue , 
    filecolor=cyan,      
    urlcolor=magenta,
    pdfcreator = {\LaTeX\ and \flqq hyperref\frqq},
}

\RequirePackage{graphicx,amsmath,amssymb,bm}
\usepackage{bbm}
\usepackage{float}
\usepackage{framed} 
\usepackage{amsthm}
\usepackage{caption}
\usepackage{subcaption}
\usepackage[normalem]{ulem}
\usepackage{comment}
\usepackage{slashed} 
\usepackage{upgreek}
\usepackage{xcolor}

\graphicspath{ {./Figs/} } 

\newcommand{\blue}{\textcolor{blue}}

\newcommand{\tr}{\text{tr}}

\newcommand{\La}{{\cal{L}}}
\newcommand{\Da}{{\cal{D}}}

\newcommand{\To}{{\cal{T}}}
\newcommand{\Z}{{\cal{Z}}}

\newcommand{\bk}{\bm{k}} 
\newcommand{\pole}{\text{pole}}
\newcommand{\zero}{\text{zero}}
\newcommand{\dslash}{\slashed{D}}
\newcommand{\pslash}{\slashed{\partial}}

\def\be{\begin{equation}}
\def\ee{\end{equation}} 
\def\bsh{\begin{shaded}}
\def\esh{\end{shaded}} 
\def\bpm{\begin{pmatrix}}
\def\epm{\end{pmatrix}}

\begin{document}
\title{Global anomalies of Green's function zeros} 
\author{Lei Su}
\affiliation{Department of Physics, University of Chicago, Chicago, Illinois 60637, USA}
\author{Ivar Martin}
\affiliation{Materials Science Division, Argonne National Laboratory, Lemont, Illinois 60439, USA}
\affiliation{Department of Physics, University of Chicago, Chicago, Illinois 60637, USA}

\begin{abstract}
We study global anomalies of nonlocal effective theories proposed to describe symmetry-preserving Luttinger surfaces, i.e., the momentum-space manifolds of Green's function zeros (GFZs) at zero energy, in strongly interacting fermionic systems. In particular, we focus on simplest possible cases associated with a gapless Dirac zero, which is the counterpart of the gapless Dirac quasiparticle in weakly interacting systems. These theories may be derived by integrating out low-energy degrees of freedom that do not couple to the relevant gauge field. We discuss the global anomaly, the bulk-boundary correspondence, and the constraint on phases consistent with the anomaly, such as non-Fermi liquids and emergent gapless quasiparticles on Luttinger surfaces. Failing to avoid a spontaneous symmetry breaking in the thermodynamical limit inevitably leads unstable GFZs. We also provide some perspectives on why the related nonlocal fermionic eﬀective theory studied recently is not a suitable starting point for a symmetrically gapped phase. 
\end{abstract} 
 
\maketitle 



\section{Introduction} 
Green's functions encode important properties of quantum systems. The poles of the two-point Green's functions (or propagators) in frequency/momentum space, have traditionally been center stage since they may be interpreted as quasiparticles in a many-body system, be it free or interacting. By contrast, Green's function zeros (GFZs), where $\det{G(\omega, \bk)} =0$, are far less studied. A GFZ occurs if there is a pole in the self-energy or the renormalization factor $Z$ vanishes. In high-energy physics, $Z =0$ is usually interpreted as the existence of a composite particle \cite{weinberg1995quantum}. In condensed matter physics, the surface formed by GFZs in  momentum space at zero frequency is called the Luttinger surface and can emerge in strongly correlated systems \cite{dzyaloshinskii2003}. The violation of Luttinger's theorem, which relates the volume of the quasiparticle Fermi surface to the density of electrons, in the presence of Luttinger surfaces has been studied intensively (see, e.g. Ref.~\cite{skolimowski2022} and references therein). 

In recent years, GFZs have shown up frequently in the context of interacting symmetry-protected topological (SPT) phases \cite{chen2013} because the many-body topological invariants in some systems involve products of  $G^{-1}(\omega, \bk)\partial G(\omega, \bk)$, with $G(\omega, \bk)$ being the two-point interacting Green's function \cite{volovik2003universe, wang2010topological, gurarie2011single, essin2011bulk, wang2012simplified, you2014topological, slagle2015exotic}. See, e.g. Refs.~\cite{zhao2023failure, Gavensky2023, blason2023, setty2023electronic,  bollmann2023topological, chen2024dirac}  for some recent developments in this direction. The bulk-boundary correspondence was proposed to be generalizable to zeros in SPT phases \cite{gurarie2011single, manmana2012, wagner2023mott, wagner2023edge}.  Another context in which GFZs have been studied is the so-called symmetric mass generation (SMG) problem: whether a spectral gap can be generated purely due to interaction even when bilinear mass terms are forbidden by symmetries \cite{wang2022symmetric}. This research was partially motivated by the desire to put a chiral gauge theory on a lattice and is closely related to the boundary states of SPT phases. It has been suggested that GFZs are a key diagnostic of the SMG phase \cite{wang2022symmetric}. However, a recent anomaly analysis based on an effective theory of GFZs seems to suggest that the gapped fermions due to SMG retain some residual signature in the infrared (IR), thus preventing the construction of a microscopic ultraviolet (UV) chiral gauge model \cite{golterman2024propagator}.

The 't Hooft anomaly analysis has played an important role in both high energy physics (see, e.g., Ref.~\cite{weinberg1996quantum})  and condensed matter physics (see, e.g. Ref.~\cite{witten2016fermion}). A global symmetry which cannot be gauged consistently is said to have a 't Hooft anomaly \cite{hooft1980naturalness}.  The existence of a nontrivial 't Hooft anomaly for a global symmetry $\cal G$ rules out a trivially (nondegenerate, symmetry-preserving) gapped ground state. It is known that the anomaly inflow mechanism is a very natural way to explain the bulk-boundary correspondence \cite{callan1985anomalies}: In the context of SPT phases, the anomaly on the boundary is compensated by that in the bulk.

To explain the puzzle from Ref.~\cite{golterman2024propagator} and to draw connections to the bulk-boundary correspondence for GFZs, we investigate nonlocal effective theories proposed to describe a GFZ. For a given propagator $G$ of a fermionic field $\psi$, we may write the effective theory at the tree level as $\La_{\text{eff}} \sim \bar{\psi} G^{-1} \psi$ with the partition function $\Z\sim \det(G^{-1})$. If the GFZ comes from the factor $i\pslash$, the partition function, at the tree level, is $\Z  \propto \det(i\pslash)^{-1}$, which is the reciprocal of the partition function of a massless Dirac fermion  $\det(i\pslash)$, up to some trivial gapped systems. Since a trivially gapped system is anomaly-free, the anomaly in the system with such a GFZ can be canceled by that of a massless Dirac fermion. This is the intuition of the global anomaly of GFZs we will use in our work. 

In the following, we analyze global anomalies in nonlocal effective theories describing a ``Dirac zero," and discuss the bulk-boundary correspondence and the constraint on ground state phases in those theories, extending the analysis of Ref.~\cite{witten2016fermion} from Dirac poles to Dirac zeros. We highlight the risk of using such nonlocal effective theories and  suggest that the analysis in Ref.~\cite{golterman2024propagator} is not a good starting point for SMG.

\section{Effective theory}
In a many-body system of fermions, we can define the time-ordered two-point Green's function, i.e., the propagator
$  
G_{ab}(x; x')  = -i \langle \To \psi_a(x) \bar{\psi}_b (x')\rangle
$, where $x = (t, \bm{x})$, and $a$ and $b$ are flavor indices, with the average taken with respect to the ground state. With space and time translational invariance, the Green's function in the frequency/momentum space  takes the general form 
 $  G(k)  = 1/[\slashed{k} -\Sigma(\slashed{k})],$
which follows from the Dyson equation for the Dirac fermions. Here $k = (\omega, \bk)$ and $\slashed{k} = \sum_{\mu =0}^d \gamma^{\mu} k_{\mu}$ with gamma matrices $\gamma^{\mu}$ satisfying $\{\gamma^{\mu}, \gamma^{\nu}\} = 2\eta^{\mu\nu}$. The (mostly plus) signature $\eta^{\mu\nu} =(-1, 1,..., 1)$ is used. We have omitted the flavor indices since we will mostly consider a single flavor. The poles of the Green's function can be interpreted as quasiparticles, which play an important role in thermal and transport properties.  In an interacting system, there can also be GFZs,  where $\det{G(\omega, \bk)} =0$. In particular, the roots $\bk$ of the GFZs at zero energy define the Luttinger surface. In the following, we will work in the relativistic framework and focus on the simplest effective theory for GFZs of the Dirac type:
\be 
G(k) = \frac{\slashed{k}}{k^2 + m^2 -i\epsilon},
\label{eq:gfz}
\ee   
with infinitesimal $\epsilon>0$. This occurs when $\Sigma(k) = -m^2/\slashed{k}$, and then the Luttinger surface is a single point.

There are different ways to obtain the low energy effective theory of a system. If we know the UV Lagrangian, it is typical to write down an effective action of a system by integrating out heavy modes. This is the top-down approach \cite{schwartz2013quantum}. In this process, nonlocal terms may be introduced but nevertheless can be viewed as a local theory by expanding them with respect to the mass of heavy modes. For example, consider a massless Dirac fermion $\psi$ coupled to a massive real scalar field $\phi$ of mass $M$ with a Yukawa coupling strength $g$ with Lagrangian
$ 
\La = \bar{\psi} i\pslash \psi + \frac{1}{2}\partial^{\mu} \phi \partial_{\mu} \phi + \frac{1}{2} M^2\phi^2 + g \phi \bar{\psi}\psi.  
$ 
Since the Lagrangian is quadratic in $\phi$, we can integrate out the scalar field exactly and obtain the nonlocal term $\frac{1}{2} g^2 \bar{\psi}\psi \frac{1}{\partial^2 +M^2} \bar{\psi}\psi$. A local effective theory is obtained by expanding it with respect to $M$ and keeping a finite number of terms. But the scheme fails when $M^2 =0$ and the resulting theory becomes nonlocal. Similarly, we can consider two Dirac fermions in odd spacetime dimensions
\be 
\La = \bar{\psi} i\dslash \psi + \bar{\psi}' i\dslash \psi'   + i m \bar{\psi}' \psi - i m\bar{\psi} \psi',  
\label{eq:2fermions}
\ee 
where $D_{\mu} = \partial_{\mu} + iA_{\mu}$.
This Lagrangian is invariant under the $U(1)$ rotation and the antiunitary time reversal operation: $T\psi(t, \bm{x}) = \gamma_0 \psi(-t, \bm{x})$ and $T\psi'(t, \bm{x}) = \gamma_0 \psi'(-t, \bm{x})$. 
We can integrate out $\psi'$ and $\bar{\psi}'$ and obtain
\be 
\La_{\text{eff}} = \bar{\psi} i\dslash  \psi + m^2 \bar{\psi}  \frac{1}{ i\dslash} \psi,
\label{eq:1fermioneff}
\ee 
which apparently is also nonlocal. It is known that the Lagrangian in Eq.~(\ref{eq:2fermions}) is anomaly-free since the system is fully gapped. However, in a recent study \cite{golterman2024propagator}, the authors started with a nonlocal effective theory similar to Eq.~(\ref{eq:1fermioneff}) and found a perturbative anomaly. We will discuss the subtle reason for this below. 

In general, the UV Lagrangian may be unknown; then we may take the bottom-up approach and write down the effective action that observes the symmetry requirement and captures the low-energy dynamics, i.e. matching all correlators up to certain orders. Given the propagator $G(x; x')$, we may write down the simplest effective theory that captures the behavior at the tree level $  
\La_{\text{eff}} = \bar{\psi} (x) G^{-1}(x; x') \psi(x').
$ 
It belongs to an infinite family of Lagrangians that yield the same propagator at the tree level. In particular, we refer to the Lagrangian in Eq.~(\ref{eq:1fermioneff}) for a single flavor as a single gapless Dirac zero.

As we mentioned earlier, the associated $\La_{\text{eff}}$ is  nonlocal in general. Nonlocality is sometimes associated with nonunitarity of the system. The reason is that the states integrated out are missing from the Hilbert space \cite{schwartz2013quantum}. Even though $\psi$ and $\psi'$ are on an equal footing in Eq.~(\ref{eq:2fermions}), to obtain an effective theory for $\psi$, $\psi'$ has been integrated out. We will consider only the case where the missing states are not charged under the gauge field associated with the anomaly or at least do not contribute extra terms to the anomaly. In other words, \textit{the Green's functions of the form in Eq.~(\ref{eq:gfz}) are associated only with degrees of freedom charged under relevant symmetries, and the GFZs refer to zeros in this subspace.}  This is crucial in our analysis. For example, as we will see, the Lagrangian in Eq.~(\ref{eq:1fermioneff}) appears to carry a global anomaly under a large $U(1)$ gauge transformation even though the complete  Lagrangian in Eq.~(\ref{eq:2fermions})  does not. The reason is  that both $\psi$ and $\psi'$ are charged under $U(1)$. Thus, while we can integrate out $\psi'$ and consider a GFZ  associated with only $\psi$, we cannot ignore the contribution to the partition function generated by integrating out $\psi'$, which restores gauge invariance. On the other hand, if $\psi'$ were neutral under $U(1)$, we could focus on the GFZ associated with only $\psi$. Clearly, the Lagrangian in Eq.~(\ref{eq:2fermions}) does not allow for $U(1)$ symmetry acting solely within the $\psi$ sector. A quantum mechanical example in which a targeted action of $U(1)$ symmetry on just one flavor of fermions is possible is presented in the Appendix.

If the system has global symmetries, the form of the Green's functions and hence  of the effective Lagrangian is constrained. It is easy to show that if $ 
\La  = \bar{\psi}(x) G^{-1}(x; x') \psi(x') $ has a global symmetry $\cal G$, which we assume, for simplicity, includes only  internal symmetries such as the global $U(1)$ of the particle number and time reversal $T$, so does $\La'  = m^2 \bar{\psi} (x) G(x; x') \psi(x')$. Here $m$ has the dimension of mass. A direct corollary is that the Lagrangian in Eq.~(\ref{eq:1fermioneff}) has $U(1)$ and $T$. Note that the first term corresponds to a pole of the Green's function $G(k)  = 1/(\slashed{k} -i\epsilon)$, which describes a single gapless Dirac fermion on the surface of a time reversal invariant  topological insulator with symmetry $U(1)$ and $T$ in $(3+1)d$ \cite{chiu2016}. A single gapless Dirac zero in Eq.~(\ref{eq:1fermioneff}) is then in the same symmetry class.

\section{Global anomalies of Dirac zeros}  
Our primary interest is 
the anomaly associated with a single gapless Dirac zero whose propagator is given by Eq.~(\ref{eq:gfz}). Our analysis parallels that of Ref.~\cite{witten2016fermion}, in which the author studied the global anomaly and the bulk-boundary correspondence of massless Dirac fermions, manifested by the pole of the propagator; thus we briefly recap it here 
\footnote{The discussion in Ref.~\cite{witten2016fermion} is divided into three scenarios: pseudoreal fermions, real fermions, and complex fermions, depending on the representation of the rotation group in the Euclidean signature, together with other symmetries. We are discussing the pseudoreal case corresponding to the boundary of a topological Mott insulator in $(3+1)d$.}. 

The partition function of a massless Dirac fermion in $(2+1)d$ (and in the Euclidean signature) is given by \be 
\Z_{\pole} = \det (i\dslash) = \prod_i \lambda_i,
\ee 
where $\lambda_i$ are the eigenvalues of the Hermitian Dirac operator $i\dslash$: $i\dslash \phi_i = \lambda_i \phi_i$. Thus $\lambda_i$ are real. Formally, the partition function is also real. This is required by time reversal inverse (TRI) on an orientable spacetime of Euclidean signature since time reversal changes its orientation and complex-conjugates the partition function. 

A global anomaly shows up if we want to make the partition function $\Z_{\pole}$ both real (to preserve TRI) and invariant under the large gauge transformations of the background gauge field $A$ \footnote{We ignore the gravitational field here for simplicity.}.  Indeed, let us start with an arbitrary $A = A_0$ and let $A_0^{\varphi}$ be the gauge field of $A_0$ after a large gauge transformation $\varphi$. If we continuously tune the parameter $s$ from 0 to 1 in
$A_s = (1-s)A_0 + s A_0^{\varphi}$,
$\Z_{\pole}$ changes continuously as $A_s$ varies from $A_0$ to $A_0^{\varphi}$. As $s$ changes, its sign is flipped every time a (real) eigenvalue $\lambda_i$ passes through zero , leading to a net spectral flow in $\Z_{\pole}$. For the partition function to be gauge
invariant under the large gauge transformation, the sign
of the partition function has to be the same. However, in the case of a massless Dirac fermion in $(2+1)d$, as the boundary theory of a $(3+1)d$ topological insulator, the partition function $\Z_{\pole}$ changes by $(-1)^n$, where $n$ is the Dirac index on a $4d$ manifold, a mapping torus \cite{witten2016fermion}. In other words, $\Z_{\pole}$ is not invariant under some large gauge transformations, giving rise to an anomaly. 

Since the infinite product in $\Z_{\pole}$ is divergent, implicit in the discussion above is that the partition function has been regularized in a way that preserves TRI.  One can also regularize it in a gauge-invariant way by introducing a massive bosonic Pauli-Villars regulator field $\chi$ that obeys the equation $(i\slashed{D} +i \mu)\chi =0 $. In this case, the partition function is no longer real.
Instead, the regularized partition function becomes
$ 
\Z_{\pole} =  \prod_i \frac{\lambda_i}{\lambda_i +i\mu},
$
which for large $\mu >0$ becomes
$ \Z_{\pole} = |\Z_{\pole}| e^{-i \frac{\pi}{2} \upeta}$, where $\upeta = \lim_{s\to 0}  \sum_i \text{sgn}(\lambda_i) |\lambda_k|^{-s}$  is a regularized difference between the number of positive and negative eigenvalues of $i\dslash$. One way to recover TRI is to add a bulk in $(3+1)d$ to cancel the anomaly through the anomaly inflow mechanism, a manifestation of the bulk-boundary correspondence. This is also the mathematical content of the Atiyah-Patodi-Singer (APS) index theorem. For more details, we refer the reader to Ref.~\cite{witten2016fermion}.

The discussion of the anomaly of Dirac zeros in Eq.~(\ref{eq:1fermioneff}) is similar. The corresponding partition function is given by
\be \Z_{\zero} = \det (i\dslash + \frac{m^2}{i\dslash}) = \prod_i \frac{\lambda_i^2 +m^2}{\lambda_i}.
\label{eq:Zpole}
\ee   
There is a simple relation between the two partition functions: $\Z_{\pole} \Z_{\zero} = \prod_i (\lambda_i^2 +m^2)$. As the product is formally positive, it is natural to expect that the anomaly in $\Z_{\zero}$ is actually canceled by $\Z_{\pole}$. A subtlety is that the regularized $\Z_{\pole}$ can vanish, leading to a divergence in $\Z_{\zero}$. We may regard this reciprocal behavior as two equivalent interpretations of the spectral inflow: Since we care about only the relative change in the number of positive and negative eigenvalues (in the infinite tower), we can regard an eigenvalue of $i\dslash$ passing through zero as an eigenvalue passing through the UV cutoff. This IR/UV dual interpretation is a common feature in discussing anomalies \cite{shifman2022advanced}. However, we can also construct a two-pole variant of the simplest Lagrangian that implements the same effect while staying away from  infinities. 

\section{Two-pole variant}
The simple two-pole variant of the Lagrangian in Eq.~(\ref{eq:1fermioneff}) can be generated by the partition function 
\be 
 \Z  = \frac{1}{2}[\Z'(m) + \Z'(-m)]  = \frac{1}{2}[\det(\dslash  -m)  +\det(\dslash+m)],
 \label{eq:2pole}
 \ee
where $ \Z'(\pm m) = \int \Da \bar{\psi} \Da \psi e^{- 
\int   \bar{\psi} (i\dslash \pm m )\psi}$ describes fermions with masses of opposite signs. When $m$ vanishes, the total partition function reduces to that of a massless Dirac fermion. For $m\neq 0$, we can obtain, by combining two additive integrands into one, the effective action $  S_{\text{eff}} = - \ln(e^{- 
\int   \bar{\psi} (i\dslash + m )\psi}+ e^{- 
\int   \bar{\psi} (i\dslash - m )\psi}),$
which is generically nonlocal for $m\neq 0$. Importantly, the partition function in Eq.~(\ref{eq:2pole}) is invariant under $T$ and $U(1)$ and produces the same two-point Green's function [Eq.~(\ref{eq:gfz})] when the background gauge field $A = 0$. The effective Lagrangian of the two-pole variant is different from the previous one in Eq.~(\ref{eq:1fermioneff}) in nonlocal terms. 

Note that if $m$ had the meaning of a scalar field, it would be natural for the system to undergo a spontaneous symmetry breaking (SSB). Unbalancing terms with $\pm m$ by an SSB would make the GFZ disappear. To stabilize the GFZ, we must assume that there are additional dynamical terms that keep $m$ disordered, fluctuating between $+m$ and $-m$. This is similar to the fluctuating order parameter approach to constructing GFZ \cite{you2018symmetric, wang2022symmetric}. These additional dynamics-generating terms should not affect the global anomaly even though they can distort the Green's function. Alternatively, we may treat $m$ as an effective \textit{parameter}, which does not have the interpretation of an order parameter.  In the Appendix, we present a quantum mechanical example in which a partition function of a similar form is naturally produced. There, the partition function $\Z$ (without the external gauge field) can be interpreted as a Dirac fermion coupled to an effective Ising degree of freedom with strength $m$.

When the external gauge field is turned on, both $\Z'(\pm m)$ can be separately regularized in a gauge-invariant way under $U(1)$ gauge transformations. However, there is no regularization scheme such that TRI is preserved simultaneously. In fact, integrating the fermions out in the large $m$ limit gives a Chern-Simons term with level $\text{sgn}(m)/2$. We cannot add a Chern-Simons counterterm to the total partition function in Eq.~(\ref{eq:2pole}) without breaking $T$ since it changes the sign of the level as well. For the half-integer level $\text{sgn}(m)/2$, the sign of the partition function is flipped after a large gauge transformation $\varphi$ under which one eigenvalue $\lambda_i$ crosses zero. Moreover, unlike Eq.~(\ref{eq:Zpole}),  both regularized $\Z'(\pm m)$ are finite and do not diverge as $A_s$ is continuously varied. This change in behavior is due to higher order terms in $\psi$ and $\bar{\psi}$ in $S_{\text{eff}}$.

\section{Bulk-boundary correspondence}
Similar to the case of gapless Dirac poles, the anomaly of gapless Dirac zeros and their variants can be compensated by a higher-dimensional bulk. Since $\Z_{\pole} \Z_{\zero}$ is anomaly-free, the sign change in $\Z_{\zero}$ after a large gauge transformation, same as that of $\Z_{\pole}$, is equal to the Dirac index in the $(3+1)d$ bulk associated with the ordinary Dirac operator. By the index theorem, this index is also directly related to the topological action describing the instanton number of the $U(1)$ field $I = \theta P = \theta (e^2/32\pi^2) \int \epsilon^{\mu\nu\alpha\beta} F_{\mu\nu}F_{\alpha\beta}$ with $\theta = \pi$, where $F_{\mu\nu} = \partial_{\mu} A_{\nu} - \partial_{\nu} A_{\mu}$.     Based on the two-pole variant of the Lagrangian, it is natural to speculate that such a local action $I$ describes the response of the bulk to the $U(1)$ field. Since $P$ is an integer on a compact surface, $\theta$ is identified with $\theta +2\pi$. There is a clear physical interpretation of $\theta = \pi$. We take the topological insulator to live in a $3d$ ball with surface $S^2$. When a unit magnetic monopole tunnels into the ball, a unit flux passes through the surface and induces a half-integer charge on the surface. The monopole itself also carries a half-integer charge through the Witten effect \cite{rosenberg2010}. This half-integer charge should be present even in the case of GFZs.

Note that we assumed that the GFZ is captured by Eq.~(\ref{eq:1fermioneff}) or its variations. If it represents the boundary of a topological Mott insulator (TMI) captured by the action above, the edge state can be gapped out completely by  putting it in contact with the usual gapless Dirac fermion on the boundary of the noninteracting topological insulator  because the combined system is anomaly-free. Naturally, the edge state should have a nontrivial response to the $U(1)$ electromagnetic field. This is compatible with the observation in the $(1+1)d$ case studied in Ref.~\cite{wagner2023mott}. Some evidence of annihilation of zeros of the TMI and poles of the noninteracting topological insulator in the $(2+1)d$ case was also presented in Ref.~\cite{wagner2023mott}.  We note that the bulk-boundary correspondence in a different model of the $(2+1)d$ TMI presented in Ref.~\cite{wagner2023edge} was interpreted in terms of charge-neutral excitations. Our analysis implies that the latter example of the TMI has to be topologically trivial with respect to the usual electromagnetic vector potential $A$ while being nontrivial with respect to the fictitious gauge field associated with the conservation of the emergent {\em neutral} quasiparticles;  the bulk-boundary correspondence is then associated with the latter. In this case, the charged degrees of freedom at the interface between the TMI and the noninteracting topological insulator remain gapless. Further work is needed to ascertain whether this is indeed the case.

\section{Constraints on the phases}
There is a family of effective theories that gives the same Dirac GFZ at the tree level. These theories may have different ground states.
As mentioned above, if we start with the effective Lagrangian in Eq.~(\ref{eq:2pole}), the ground state is likely to be in the SSB phase where $m$ plays the role of an order parameter. Then the Dirac zero is unstable and the  ground state is degenerate. The  stabilization of anomalous GFZs would require a special mechanism, such as order parameter fluctuations \cite{you2018symmetric, wang2022symmetric}.  Assuming that there is no extra contribution to the anomaly, we can constrain the possible phases of these Lagrangians if there is no such SSB.

The fermionic field $\psi$ has gapped poles at $\pm m$, but the possibility that an emergent gapless quasiparticle at the Luttinger surface exists while both $U(1)$ and $T$ are preserved is not ruled out. It was proposed recently that such an emergent quasiparticle exists on the Luttinger surface \cite{fabrizio2022emergent, fabrizio2023spin}, and a bulk-boundary correspondence of the emergent neutral excitations was discussed in relation to the TMI \cite{wagner2023edge}. It is clear from our analysis  that a neutral quasiparticle cannot match the anomaly of a single Dirac zero. If the emergent quasiparticles account for the anomaly, they must be charged under the $U(1)$ gauge field even though they do not contribute to the particle density. Thus an emergent particle can be, e.g., a composite particle of a fermion and a neutral boson or a fractionalized particle such as a fermionic holon. 

Another option is a non-Fermi liquid. Since GFZs exist only in interacting systems, strong correlations can lead to a breakdown in  the conventional Landau Fermi liquid quasiparticle picture. In the IR, the Green's function is scale invariant and the ground state can be conformally invariant. It may well correspond to an ``un-Fermi liquid" \cite{phillips2013} based on ``unparticles" proposed in Ref.~\cite{Georgi2007}. Some effective fields arising from integrating out the UV degrees of freedom couple to unparticle stuff and lead to GFZs. $m$ in Eq.~(\ref{eq:2pole}) may be viewed as such a field \footnote{Note that instead of integrating over $m$ in the Lagrangian as in Ref.~\cite{phillips2013}, our two-pole variant shows that the ``unparticle'' comes from a sum/integral in the partition function over $m$.}. Unlike the first term, the second term in the Lagrangian in Eq.~(\ref{eq:1fermioneff}) is nonlinear in the gauge field $A$, and the conventional expression for the  current breaks down.  The special property of the (incoherent) unparticle field is that it still has to carry the current in response to the gauge field even though it does not contribute to the particle density of $\psi$ (since the renormalization factor $Z$ vanishes).

\section{Anomaly-free case}
Let us now discuss the anomaly-free case in order to illustrate the assumption that the missing states or states integrated out should not contribute to the associated 't Hooft anomaly. We start with  two Dirac fermions with TRI masses as in Eq.~(\ref{eq:2fermions}). The partition function (in the Euclidean signature) is given by 
\begin{align}
 \Z &= \int \Da \bar{\psi} \Da \psi \Da \bar{\psi}' \Da \psi'  e^{-\int \La}   =  \det(i\dslash) \int \Da \bar{\psi} \Da \psi e^{-
\int \La_{\text{eff}} } \nonumber \\
&= \det(i\dslash)  \det(i\dslash + \frac{m^2}{ i\dslash})  = \det (D^2 +m^2) ,    
\end{align}
where $\La_{\text{eff}}$ is exactly the same as Eq.~(\ref{eq:1fermioneff}). Therefore, the total partition function is free from the global anomaly since both fermions are gapped.  However, if we start with $\int \Da \bar{\psi} \Da \psi \exp (- 
\int   \La_{\text{eff}})$ directly,  without the contribution from the  factor $\det(i\dslash)$, there is a global anomaly. 

Even though we have been focusing on global anomalies, it is easy to see that the observation also has implications for perturbative anomalies. In particular, Ref.~\cite{golterman2024propagator} starts with an effective Lagrangian similar to Eq.~(\ref{eq:1fermioneff}) and explicitly shows the existence of axial anomalies in the context of SMG. This clearly creates a puzzle; that is, if a fermion system with no anomaly has been gapped out by interactions, how can there still be nontrivial anomalies in the IR? The observation above reveals a simple solution to the puzzle: in the SMG case, an effective Lagrangian similar to Eq.~(\ref{eq:1fermioneff}) can not capture all the low energy physics. For example, if eight flavors of Majorana fermions at the end of the Fidkowski-Kitaev model \cite{fidkowski2010, fidkowski2011} can be gapped out via SMG, the anomaly can not correspond to a single flavor with a gapless GFZ. Proper accounting of the integrated out states would cure the apparent anomaly mismatch coming from a single flavor. Note that we can rewrite Eq.~(\ref{eq:1fermioneff}) as
$ 
\La_{\text{eff}} = \det(i\dslash)^{-1} \Z,
$ 
and represent $\det(i\dslash)^{-1} $ as a path integral in terms of bosonic $\chi$ and $\bar{\chi}$: 
\be   
\det(i\dslash)^{-1} = \int \Da \bar{\chi} \Da \chi e^{- 
\int   \bar{\chi} i\dslash   \chi}. 
\ee 
Since $\Z$ is anomaly-free, such bosonic fields provide an effective way to calculate the anomaly in $\La_{\text{eff}}$, consistent with the observation in  Ref.~\cite{golterman2024propagator}.

\section{Conclusion}
We extended the anomaly analysis to  systems with a Luttinger surface described by nonlocal effective Lagrangians, under the assumption that the states integrated out do not contribute additional terms to the anomaly. In particular, we have argued that the nonlocal Lagrangian in Ref.~\cite{golterman2024propagator} does not describe a symmetrically gapped phase since it violates this assumption.
Our focus has been on a single gapless Dirac zero in $(2+1)d$ protected by $U(1)$ and time reversal $T$, though the analysis can be generalized \cite{witten2016fermion}. Note that our discussion is based on the assumption of existence of a nonlocal effective theory that yields a gapless GFZ. From the two-pole invariant in Eq.~(\ref{eq:2pole}), an SSB in $m$ would destabilize the GFZ. It is conceivable that such an SSB in the thermodynamical limit is inevitable, making it impossible to construct a stable gapless Dirac zero on the boundary if the bulk is topologically nontrivial with respect to the gauge field associated with the charge. To demonstrate the existence of a stable Dirac zero, one would need to construct phases that avoid SSBs using microscopic or field-theoretical models \footnote{However, for Luttinger surfaces where Luttinger's theorem holds \cite{heath2020necessary}, the 't Hooft anomaly should play an important role, just as it does for Fermi surfaces \cite{else2021}.}. The study of gapless GFZs can provide new insights into strongly correlated systems. 
 
\textit{Note added.} Recently, we became aware of a related work \cite{zeng2024}. 

\acknowledgments
 
We thank M. Zeng for discussions. This work was supported by the US Department of Energy, Office of Science, Basic Energy Sciences, Materials Sciences and Engineering Division.  
 
\appendix
\section{Global anomaly in the single-site Hubbard model}
\label{sec:qm}
As an illustration of the anomaly analysis in the context of GFZs, we study the global anomaly in the quantum mechanics of a single-site Hubbard model by generalizing the global anomaly of a single massless Dirac fermion protected by charge-conjugation and global $U(1)$ \cite{Elitzur1986}.  This is also a toy model for the two-pole representation in the main text.

\subsection{Toy model}
Consider the single-site Hubbard model described by the  Hamiltonian 
\be 
H =   U (c^{\dagger}_{\uparrow}c_{\uparrow} + c^{\dagger}_{\downarrow}c_{\downarrow} -1)^2 = U(2 n_{\uparrow} n_{\downarrow} - n_{\uparrow} -n_{\downarrow} +1).
\ee 
It is easy to see that this Hamiltonian is invariant under the charge conjugation $ 
{\cal{C}} c_s {\cal{C}}^{-1} = c_s^{\dagger}$ for $s = \uparrow/\downarrow$.
Essentially, the reason is that $N_{s} \equiv  n_s -1/2 = c^{\dagger}_{s} c_{s}  -1/2$ with $s =  \uparrow/\downarrow $ changes  sign under charge conjugation. This toy Hamiltonian belongs to a class of particle-hole symmetric Hubbard models on a bipartite lattice \cite{chiu2016}. 

In the four-dimensional Hilbert space, the ground states span a two-dimensional subspace   with the basis states $|n_{\uparrow} =1, n_{\downarrow} = 0\rangle $ and $|n_{\uparrow} =0, n_{\downarrow} = 1\rangle $. Since $n_{\uparrow}$ and $n_{\downarrow}$ are separately conserved in the Hamiltonian, there are two independent  global $U(1)$ symmetries. We can see that charge conjugation $\cal{C}$ and one of the $U(1)$ symmetries, say, $U(1)_{\uparrow}$, in general are not simultaneously preserved in the ground state. Indeed, $\frac{1}{\sqrt{2}}(|n_{\uparrow} =1, n_{\downarrow} = 0\rangle  \pm |n_{\uparrow} =0, n_{\downarrow} = 1\rangle)$ preserves the charge conjugation but breaks $U(1)_{\uparrow}$. As we will see, this is due to the global anomaly between charge conjugation and either one of the $U(1)$ symmetries. Other symmetry-preserving deformations cannot change this conclusion. Note, however, that there is no anomaly between charge conjugation and the diagonal $U(1)$, i.e. the global $U(1)$ symmetry associated with $n_{\uparrow} + n_{\downarrow}$. There are other symmetries we can consider for this toy model, see, e.g. Ref.~\cite{manmana2012}. 
 
To obtain an effective theory, we can consider the partition function $\Z = \tr(e^{-\beta H})$ after tracing over the spin down sector where $n_{\downarrow} =0, 1$: 
\begin{align}
 \Z &= \tr[e^{-\beta U (-n_{\uparrow} +1)}] + \tr[e^{-\beta U n_{\uparrow}}] \nonumber \\ 
 &= e^{-\beta \frac{U}{2}}\left(\tr[e^{-\beta (-U)(n_{\uparrow} -\frac{1}{2})}] + \tr[e^{-\beta U (n_{\uparrow} - \frac{1}{2})}]\right) \nonumber \\ 
 & =e^{-\beta \frac{U}{2}}\int {\mathcal{D}}\bar{\psi} {\mathcal{D}} \psi [  e^{-\beta \frac{U}{2}}e^{- \int_0^{\beta}  d\tau  \bar{\psi}((1+\epsilon U) \partial_{\tau} -U) \psi }  \nonumber \\
 &+ e^{\beta \frac{U}{2}}  e^{- \int_0^{\beta} d\tau \bar{\psi}( (1-\epsilon U) \partial_{\tau} +U) \psi } ]\nonumber \\ 
 & =e^{-\beta \frac{U}{2}}\int \prod {\mathcal{D}}\bar{\psi}_m {\mathcal{D}} \psi_{-m} [   e^{- \sum  \bar{\psi}_n  ( i \omega_n +U) \psi_{-n} }\nonumber \\
 &+ e^{- \sum  \bar{\psi}_n  ( i \omega_n -U) \psi_{-n} }  ] \nonumber \\ 
 & = e^{-\beta \frac{U}{2}}\left[ (e^{\beta \frac{U}{2}} +e^{-\beta \frac{U}{2}})  + (e^{\beta \frac{U}{2}} +e^{-\beta \frac{U}{2}}) \right] \nonumber\\
 & = 2(e^{-\beta U} +1).   
\end{align} 
Here, $\beta = 1/T$ is inverse temperature, $\psi$ and $\bar{\psi}$ are Grassmann numbers (with the spin index dropped),  $\epsilon$ is a positive infinitesimal, and antiperiodic boundary conditions along $\tau$ are used. $ \omega_n = 2\pi \beta (n+1/2)$ is the Matsubara frequency. $\psi_m$ and $\bar{\psi}_m$ are the  Fourier transforms of  $\psi(\tau)$ \cite{nakahara2018geometry}. This expression can be obtained directly by summing over the states in the Hilbert space, but we derived it using the path integral formalism to exemplify the two-pole variant in the main text:
the partition function can be written as \be
\Z = e^{-\beta \frac{U}{2}} [ \det(\partial_{\tau} -U)  +\det(\partial_{\tau} +U) ],
\ee 
which should be compared with Eq.~(6) in the main text. It is easy to check that this partition function is invariant under charge conjugation. The single-particle Green's function has a (gapless) zero: 
\be 
G (i\omega_n) = \frac{1}{2}\left(\frac{1}{i\omega_n +U} + \frac{1}{i\omega_n -U} \right)=\frac{i\omega_n}{(i\omega_n)^2 -U^2}.
\ee  
Note that in the derivation above, it is assumed implicitly that charge conjugation is not broken because otherwise we do not have to sum over  $n_{\downarrow} =0, 1$.

To obtain an effective action, we need to combine two integrands into one:
\begin{align}
\Z & \propto \int \prod {\mathcal{D}}\bar{\psi}_{-m}{\mathcal{D}} \psi_m \nonumber \\
& \times e^{- \sum \bar{\psi}_{-n}i\omega_n \psi_n + \ln (e ^{- U\sum \bar{\psi}_{-n} \psi_n}+e ^{+U \sum \bar{\psi}_{-n}\psi_n}) }\nonumber \\
& = \int \prod {\mathcal{D}}\bar{\psi}_{-m}{\mathcal{D}} \psi_m e^{- \sum  i\omega_n \bar{\psi}_{-n}\psi_n +  \frac{U^2}{2} (\sum \bar{\psi}_{-n}\psi_n)^2 }.
\end{align} 
Small $U$ is assumed in the second line. This effective action is highly nonlocal because of the interaction $\frac{U^2}{2} (\sum \bar{\psi}_{-n}\psi_n)^2$. We may make further approximations to replace the quartic term with a quadratic one. However, if we are interested only in reproducing the 
two-point Green's functions at the tree level, we may simply take 
\be 
S_{\text{eff}} = \sum \bar{\psi}_n G_n^{-1} \psi_n = \sum \bar{\psi}_n \frac{(i\omega_n)^2 -U^2}{i\omega_n} \psi_n,
\ee  
whose Lagrangian (in imaginary time) is
 \be 
\La_{\text{eff}} = \bar{\psi}  \frac{ \partial_{\tau}^2 +U^2}{\partial_{\tau} }  \psi = \bar{\psi} \left(   \partial_{\tau}  +\frac{U^2}{\partial_{\tau}}  \right)\psi.
\label{eq:qmeff}
\ee 
This is the zero dimensional analog of Eq.~(3). We  already explained in the main text that $\La_{\text{eff}}$ is invariant under  $ 
{\cal{C}} \psi {\cal{C}}^{-1} = \bar{\psi}$. In zero dimension, this can also be seen explicitly by finding the Green's function of $\partial_{\tau}^{-1}$:  
$ G(\tau, \tau') \sim \theta(\tau-\tau') -\theta(\tau'-\tau)$. Here $\theta$ is the Heaviside step function. 
 
Note that in the derivation of the effective theory above, spin-down states have been integrated out even though they are on equal footing with spin-up states. This process led to a nonlocal effective theory that is  associated with only the spin-up fermion, and the GFZ is defined in this subspace which is charged under $U(1)_{\uparrow}$. Importantly, in the process of integration, the spin-down states, which are neutral under $U(1)_{\uparrow}$,  do not make additional contributions to the anomaly associated with $U(1)_{\uparrow}$. The GFZ should be compared with zero diagonals in the Green's function associated with both fermions which can exist even if the system is noninteracting \cite{misawa2022zeros}. This point was emphasized in the main text. 

\subsection{Global anomaly}
Let us now study the global anomaly by turning on the background gauge field for $c_{\uparrow}$ associated with $U(1)_{\uparrow}$.
\be 
H = iA (c^{\dagger}_{\uparrow}c_{\uparrow}  -\frac{1}{2}) +  U (c^{\dagger}_{\uparrow}c_{\uparrow} + c^{\dagger}_{\downarrow}c_{\downarrow} -1)^2
\ee
The exact partition function (in imaginary time) is 
\be 
\Z = 2 \cos(A/ 2\beta)(e^{-\beta U }+ 1).
\ee 
The charge conjugation $\cal{C}$ acts on $A$ as $ 
{\cal{C}} A {\cal{C}}^{-1} = -A$. 
Thus the partition function is invariant under $\cal{C}$ but changes sign under the large gauge transformations $A \to A+2\pi\beta $. This is not the case if $A$ couples to both $c_{\uparrow}$ and $c_{\downarrow}$.

We can also observe the anomaly from the effective action in Eq.~(\ref{eq:qmeff}) describing one of the two species
\be 
S_{\text{eff}} =\sum_n \bar{\psi}_{-n}(i\omega_n+ i A + \frac{U^2}{i\omega_n + iA})\psi_{n},
\ee  
assuming the validity of minimal coupling. 
Under charge conjugation $ 
{\cal{C}} A {\cal{C}}^{-1} = -A$ and ${\cal{C}} \psi_{n} {\cal{C}}^{-1} = \bar{\psi}_{n}$. We regularize the system in a charge conjugation invariant way, and we have 
\begin{align}
\frac{\Z(A)}{\Z(0)} &  = \lim_{N\to \infty}  \prod_{n = -N}^{ N-1} \frac{ \frac{(i\omega_n+ i A )^2+U^2}{i\omega_n + iA}}{\frac{(i\omega_n)^2+U^2}{i\omega_n}} \nonumber \\
& = \frac{1}{ \cos(A/2\beta)} \frac{ \cos (A/\beta) +\cos(\beta U)}{1+\cos(\beta U)},
\label{eq:zz}
\end{align} 
where we have used the identity
\be 
\cos x = \prod_0^{\infty} \left( 1- \frac{x^2}{\pi^2 (n+1/2)^2}\right).
\ee 
$\Z(A)$ is no longer invariant under the large gauge transformation. Note that the factor $\cos(A/2\beta)$ is the same change in the partition function of a massless Dirac fermion after the coupling to $A$ \cite{Elitzur1986}, which is a manifestation of the fact that $\Z_{\pole} \Z_{\zero}$ is anomaly-free emphasized in the main text.

We   discussed only the simplest toy model here. Some conclusions could have been drawn without integrating out spin-down states. We used this example to substantiate the anomaly analysis in the main text where effective theories are considered. In many cases where a microscopic model is lacking, an effective theory may be helpful. The power of the anomaly analysis manifests when we consider more complicated systems  with the same anomaly. In particular, the anomaly is robust toward symmetry-preserving perturbations. Note that we may generalize the analysis to a higher-dimensional particle-hole symmetric Hubbard model on a bipartite lattice.  Another direction would be to consider Hatsugai-Kohmoto interactions, analogs of Hubbard interactions in momentum space \cite{setty2023electronic}.

\bibliography{GFZeros}

\end{document}